\documentclass[preprint,showpacs,preprintnumbers,amsmath,amssymb,aps,pre]{revtex4}

\usepackage{graphicx}
\usepackage{dcolumn}
\usepackage{bm}
\usepackage{epsfig}

\begin{document}

\title{Description of stochastic and chaotic series using visibility graphs}

\author{Lucas Lacasa, Raul Toral}
\email{{lucas,raul}@ifisc.uib-csic.es}
\affiliation{IFISC, Instituto de F\'{\i}sica Interdisciplinar y Sistemas Complejos (CSIC-UIB)\\
Campus UIB, 07122-Palma de Mallorca, Spain}%


\date{}


\begin{abstract}Nonlinear time series analysis is an active field of research that studies the structure of complex signals
in order to derive information of the process that generated those series, for understanding, modeling and forecasting purposes.
In the last years, some methods mapping time series to network representations have been proposed. The purpose is to investigate on
the properties of the series through graph theoretical tools recently developed in the core of the celebrated complex network theory.
Among some other methods, the so-called visibility algorithm has received much attention, since it has been shown that series correlations
are captured by the algorithm and translated in the associated graph, opening the possibility of building fruitful connections between time
series analysis, nonlinear dynamics, and graph theory. Here we use the \textit{horizontal visibility algorithm} to characterize and distinguish
between correlated stochastic, uncorrelated and chaotic processes. We show that in every case the series maps into a graph with exponential degree
distribution $P(k)\sim\exp(-\lambda k)$, where the value of $\lambda$ characterizes the specific process. The frontier between chaotic and
correlated stochastic processes, $\lambda=\ln(3/2)$, can be calculated exactly, and some other analytical developments confirm the results provided by extensive
numerical simulations and (short) experimental time series.
\end{abstract}

\maketitle
\textbf{Published in Physical Review E 82, 036120 (2010)}

\section{Introduction}
Concrete hot
topics in nonlinear time series analysis \cite{Libro} include the characterization of correlated stochastic processes and chaotic phenomena
in a plethora of different situations including long-range correlations in earthquake statistics \cite{terremotos1}, climate records \cite{nilo}, noncoding DNA sequences \cite{DNA}, stock market \cite{pnas_finance},
 urban growth dynamics \cite{urban}, or physiological series \cite{gold1,gold22} to cite but a few, and chaotic processes \cite{procaccia1, Farmer, sugihara, tsonis, kaplan, Libro, vulpiani}.

Both stochastic and chaotic processes share many features, and the discrimination between them is indeed very subtle.
The relevance of this problem is to determine whether the source of unpredictability (production of entropy) has its origin in
a chaotic deterministic or stochastic dynamical system, a fundamental issue for modeling and forecasting purposes. Essentially, the majority of methods \cite{Libro, vulpiani} that have been introduced so far rely on two major differences between chaotic and stochastic dynamics. The first difference is that chaotic systems have a finite dimensional attractor, whereas stochastic
processes arise from an infinite-dimensional one. Being able to reconstruct the attractor is thus a clear evidence
showing that the time series has been generated by a deterministic
system. The development of sophisticated embedding techniques \cite{Libro} for attractor reconstruction is the most representative step forward in this direction. 
The second difference is that deterministic systems evidence, as opposed to random ones, short-time prediction: the time evolution of two nearby states
will diverge exponentially
fast for chaotic ones (finite and positive Lyapunov exponents) while in the case of a stochastic process such separation is randomly distributed. Whereas some algorithms relying on the preceding concepts are nowadays available, the great majority of them are purely phenomenological and often complicated to perform,
computationally speaking. These drawbacks provide the motivation for a search for new
methods that can directly distinguish, in a reliable way, stochastic
from chaotic time series. This is, for instance, the philosophy behind a recent work by Rosso and co-workers \cite{ROSSO}, where the authors present a 2D diagram (the so-called entropy-complexity plane) that relates two information-theoretical functionals
of the time series (entropy and complexity), and compute numerically the coordinates of several chaotic and stochastic series in this plane.
The purpose of this paper is to offer a different, conceptually simple and computationally efficient method to distinguish between deterministic and stochastic dynamics.
 
The proposed method uses a new approach to time series analysis that has been developed in the last years \cite{zhang1,small,PNAS,NJPP,chinos,jstat}. In a nutshell, time series are mapped into a network representation (where the connections between nodes capture the series structure according to the mapping criteria) and graph theoretical tools are subsequently employed to characterize the properties of the series. Some methods sharing similar philosophy include recurrence networks, cycle networks, or correlation networks to cite some (see \cite{NJPP} for a comparative review).
Amongst these mappings, the so-called visibility algorithm \cite{PNAS} has received much attention, since it has been shown that series correlations (including periodicity, fractality or chaoticity) are captured by the algorithm and
translated in the associated visibility graph \cite{PNAS,EPL,PRE}, opening the possibility of building bridges between time series analysis, nonlinear dynamics, and graph theory. Accordingly, several works applying such algorithm in several contexts ranging from geophysics \cite{hurricanes} or turbulence \cite{turbulence} to physiology \cite{physio} or finance \cite{finance} have started to appear \cite{UE}.

Here we address the characterization of chaotic, uncorrelated and correlated stochastic processes, as well as the discrimination between them,
via the {\sl horizontal visibility algorithm}. We will show that a given series maps into a graph with an exponential degree distribution $P(k)\sim \exp(-\lambda k)$, where $\lambda<\ln(3/2)$
characterizes a chaotic process whereas $\lambda>\ln(3/2)$ characterizes a correlated stochastic one. The frontier $\lambda_{un}=\ln(3/2)$ corresponds to
the uncorrelated situation and can be calculated exactly \cite{PRE}, thus the method
is well grounded. Some other features are calculated analytically, confirming our numerical results obtained through extensive simulations for Gaussian fields
with long-range (power-law) and short-range (exponential) correlations and a plethora of chaotic maps (Logistic, H\'{e}non, time-delayed H\'{e}non, Lozi,
Kaplan-Yorke, $\alpha$-map, Arnold cat). Experimental (short) series of sinus rythm cardiac interbeats --which have been shown to evidence long-range correlations-- are also analyzed.
Moreover, we will also show that the method not only distinguishes but also quantifies (by means of the parameter $\lambda$) the degree of chaoticity or stochasticity of
the series. The rest of the paper is organized as follows: in section II we recall some properties of the method, and in particular we state the theorem that addresses uncorrelated
series. In section III we study how the results deviate from this theory in the presence of correlations, through a systematic analysis of long-range and short-range stochastic processes. 
Results are validated in the case of experimental time series. Similarly, in section IV we address time series generated through chaotic maps. In section V and VI analytical developments and heuristic
arguments supporting our previous findings are outlined. In section VII we comment on the current limitations of the algorithm, and in section VIII we conclude.
\section{Horizontal visibility algorithm}
\begin{figure}[h]
\centering
\includegraphics[width=0.50\textwidth]{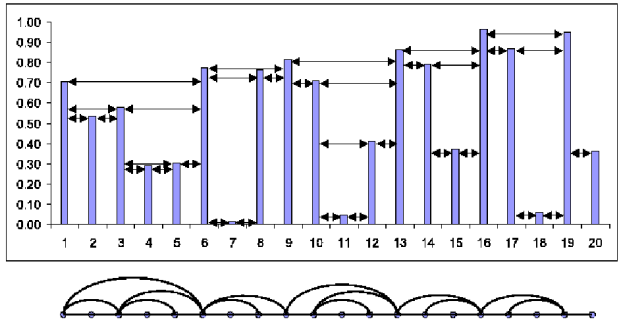}
\caption{Graphical illustration of the horizontal visibility algorithm. A time series is represented in vertical bars, and in
the bottom we plot its associated horizontal visibility graph, according to the geometrical criterion encoded in Eq. (\ref{criterio}) (see the text). \label{fig1}}
\end{figure}

The horizontal visibility algorithm has been recently introduced \cite{PRE} as a map between a time series and a
graph and it is defined as follows. Let $\{x_i\}_{i=1,. . .,N}$ be a time
series of $N$ real data. The algorithm assigns each datum of the
series to a node in the {\sl horizontal visibility graph} (HVG). Two nodes $i$ and $j$ in the graph are connected if one
can draw a horizontal line in the time series joining $x_i$ and $x_j$
that does not intersect any intermediate data height (see figure \ref{fig1} for a graphical illustration). Hence, $i$ and $j$ are two connected
nodes if the following geometrical criterion is fulfilled
within the time series:
\begin{equation}
x_i,x_j > x_n, \ \forall \ n \ \left | \ i < n < j\right. . \label{criterio}
\end{equation} 
Some properties of the HVG can be found in \cite{PRE}. Here we recall the main theorem for random uncorrelated series, whose proof can also be found in \cite{PRE}:\\

\texttt{Theorem (uncorrelated series)} Let ${x_i}$ be a bi-infinite sequence of independent and identically distributed random variables extracted from a continous probability density $f(x)$.
The degree distribution of its associated horizontal visibility graph is
\begin{equation}
P(k)=\frac{1}{3}\bigg(\frac{2}{3}\bigg)^{k-2}.
\label{pk}
\end{equation} 
Note that $P(k)$ can be trivially rewritten as $P(k)\sim \exp(-\lambda_{un} k)$ with $\lambda_{un}=\ln(3/2)$. Interestingly enough, this result is independent of
the generating probability density $f(x)$, (as long as it is a continuous one, independently on whether the support is compact or not). This result
shows that there is an universal equivalency between uncorrelated processes and $\lambda=\lambda_{un}$. In what follows we will investigate how results deviate from this
theoretical result when correlations are present.

\section{Correlated stochastic series}
\begin{figure}[h]
\centering
\includegraphics[width=0.8\textwidth]{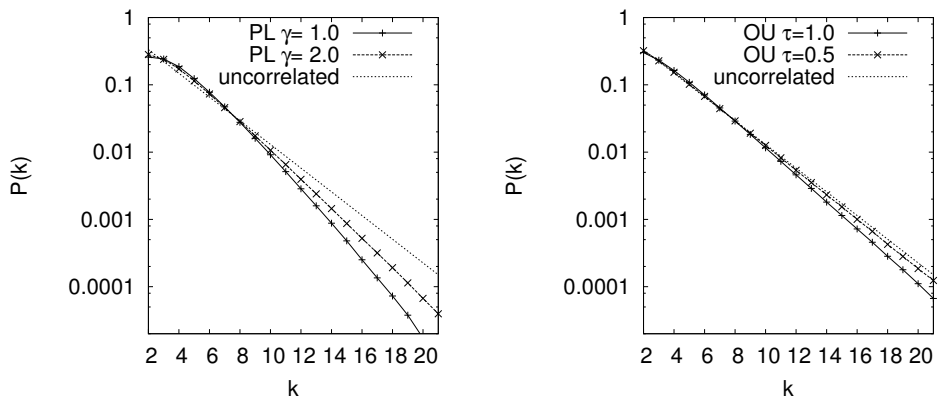}
\caption{\textit{Left}: Semilog plot of the degree distribution $P(k)$ of a Gaussian correlated series of $N=2^{18}$ data with power-law decaying correlations $C(t)\sim t^{-\gamma}$,
for $\gamma=1.0$ and $\gamma=2.0$.
showing an exponential function. $P(k)\sim \exp(-\lambda k)$ in both cases, with slope $\lambda=0.59$ and $\lambda=0.50$ respectively. For comparison, the shape of $P(k)$ associated to a random uncorrelated
series is shown, having $\lambda_{un}=\ln(3/2)<\lambda, \ \forall \gamma$. \textit{Right}: Similar results associated to short-range correlated series generated
through an Ornstein-Uhlenbeck process with correlation function $C(t)\sim \exp(-t/\tau)$. \label{fig2}}
\end{figure}

\begin{figure}[]
\centering
\includegraphics[width=1.0\textwidth]{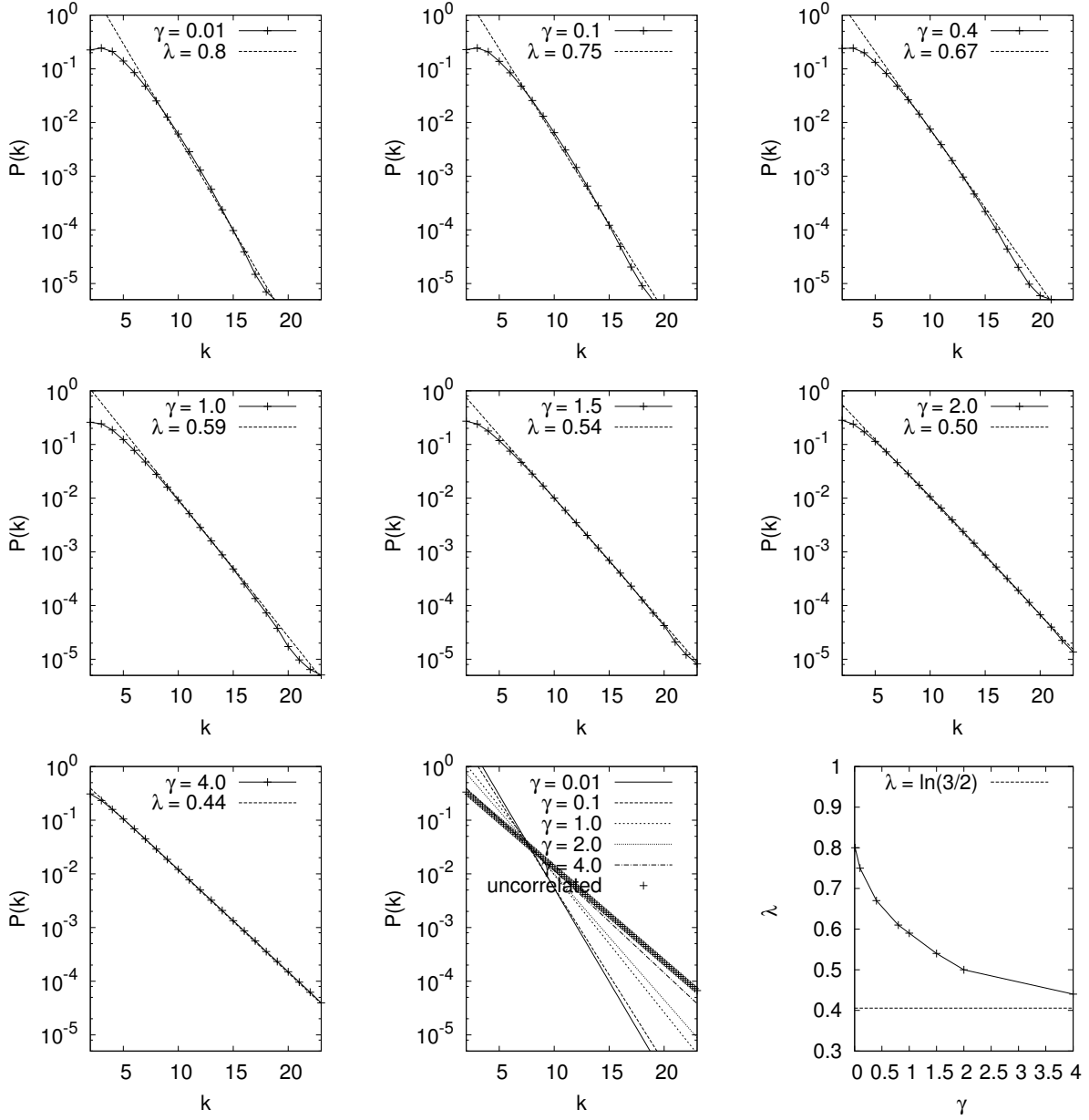}
\caption{From left to right, up to bottom: Semilog plot of the degree distributions of horizontal visibility graphs associated to long-range correlated series
with correlation function $C(t)\sim t^{-\gamma}$, for different values of $\gamma$ (data are averaged over $100$ realizations). In every case we find that the degree distribution is exponential $P(k)\sim \exp(-\lambda k)$, where the slope $\lambda$ monotonically decreases with $\gamma$. In figure 9 and 10 we plot the slope of such degree distribution for increasing
values of the correlation strength $\gamma$: the convergence towards the uncorrelated situation ($\lambda=\lambda_{un}=\ln(3/2)$) is slow, what allows us to
distinguish correlated series from uncorrelated ones even when the correlations are very weak.} \label{multi_PL}
\end{figure}

  In order to analyze the effect of correlations between the data of the series, we focus on two generic and paradigmatic
correlated stochastic processes, namely long-range (power-law decaying correlations)
and Ornstein-Uhlenbeck (short-range exponentially decaying correlations) processes. We have computed the degree distribution of the HVG associated
to different long-range and short-range correlated stochastic series (the method for generating the associated series is outlined in the next section). In the left panel of Fig.(\ref{fig2}) we plot in a semi-log scale the degree distribution for
correlated series with correlation function $C(t)=t^{-\gamma}$ for different values of the correlation strength $\gamma \in [10^{-2}-10^1]$, while in the right panel of
the same figure, we plot the results for an exponentially decaying correlation function $C(t)=\exp(-t/\tau)$. Note that in both cases the degree distribution of the
associated HVG can be fitted for large $k$ by an exponential function $\exp(-\lambda k)$. The parameter $\lambda$ depends on $\gamma$ or $\tau$ and is, in each case, a monotonic
function that reaches the asymptotic value $\lambda=\lambda_{un}= \ln(3/2)$ in the uncorrelated limit $\gamma\to\infty$ or $\tau\to 0$, respectively. Detailed results of this phenomenology can be found in
figure \ref{multi_PL}, and in the the right panel of figure \ref{diagram} where we plot the functional relation $\lambda(\gamma)$ and $\lambda(\tau)$. 
In all cases, the limit is reached from above, i.e. $\lambda>\lambda_{un}$. Interestingly enough, for the power-law correlations
the convergence is slow, and there is still a noticeable deviation from the uncorrelated case even for weak correlations ($\gamma>4.0$), whereas the convergence with $\tau$ 
is faster in the case of exponential correlations.\\

\noindent\textit{\textbf{Minimal substraction procedure\\}}
In what follows we explain the method we have used to generate series of correlated Gaussian random numbers $x_i$ of zero mean and correlation function
$\langle x_ix_j\rangle=C(|i-j|)$. The classical method for generating such correlated series is the so-called Fourier filtering method (FFM). This method
proceeds by filtering the Fourier components of an uncorrelated sequence of random numbers with a given filter (usually, a power-law function) in
order to introduce correlations among the variables.
However, the method presents the drawback of evidencing a finite cut-off in the range where the variables are actually correlated, rendering it useless in 
practical situations. An interesting improvement was introduced some years ago by Makse et. al \cite{FFM1} in order to remove such cut-off.
This improvement was based
on the removal of the singularity of the power-law correlation function $C(t)\sim t^{-\gamma}$ at $r=0$ and the associated aliasing effects
by introducing a well defined one
$C(t)=(1+t^2)^{-\gamma/2}$ and its Fourier transform in continuous-time space. Accordingly, cut-off effects were removed and variables present the desired correlations in their whole range. 

 We use here an alternative modification of the FFM that also removes undesired cut-off effects for generic correlation functions and takes in consideration the discrete nature of the series.
Our modification is based on the fact that not every function $C(t)$ can be considered to be the correlation function of a Gaussian field, since some mathematical
requirements need to be fulfilled, namely that the quadratic form $\sum_{ij} x_iC(|i-j|)x_j$ be positive definite. For instance, let us suppose that we want to
represent data with a correlation function that behaves asymptotically as $C(t)\sim t^{-\gamma}$. As this function diverges for $t\to 0$ a regularization is
needed. If we take $C(t)=(1+t^2)^{-\gamma/2}$, then the discrete Fourier transform $S(k)=N^{1/2} \sum_{j=1}^N \exp(i \frac{j k}{N})C(j)$ turns out to be negative
for some values of $k$, which is not acceptable. To overcome this problem, we introduce the \textit{minimal substraction procedure},
defining a new spectral density as $S_0(k)=S(k)-S_{min}(k)$, being $S_{min}(k)$ the minimum value of $S(k)$ and using this expression instead of the former one
in the filtering step. The only effect that the minimal substraction procedure has on the field correlations is that $C(0)$ is no longer equal to
$1$ but adopts the minimal value required to make the previous quadratic form positive definite. The modified algorithm is thus the following:
\begin{itemize}
 \item Generate a set $\{u_j\}, j = 1, . . . , N$ , of independent Gaussian variables of zero mean and variance one, and compute the
discrete Fourier transform of the sequence, $\{\hat u_k\}$.
\item Correlations are incorporated in the sequence by multiplying the new set by the desired spectral density $S(k)$, having in mind that this density
is related with the correlation function $C(r)$ through $S(k) = \sum_{r} N^{1/2}  \exp(irk) C(r)$.  Make use of $S_0(k)=S(k)-S_{min}(k)$ (minimal substraction procedure)
rather than $S(k)$ in this process. Concretely, the correlated sequence in Fourier space $\hat x_k$ is given by
$\hat x_k = N^{1/2}S_0(k)^{1/2}\hat u_k$.
\item Calculate the inverse Fourier transform of $\hat x_k$ to obtain the Gaussian field $x_j$ with the desired correlations.
\end{itemize}

\subsection{Application to real cardiac interbeat dynamics}
\begin{figure}[h]
\centering
\includegraphics[width=0.8\textwidth]{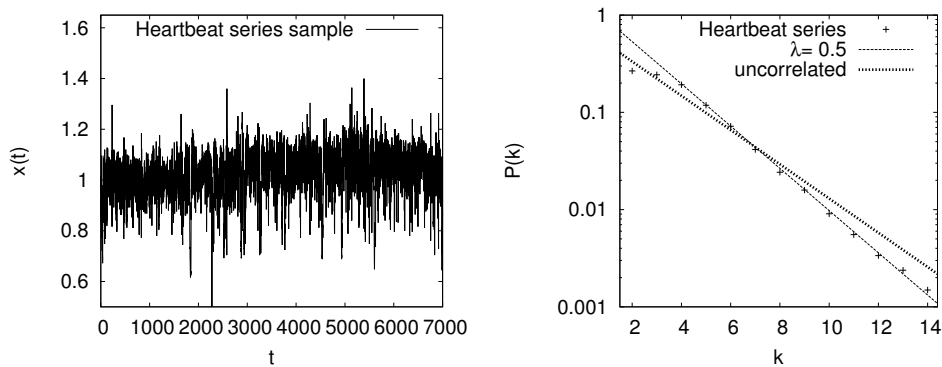}
\caption{Semi-log plot of the degree distribution of the HVG associated to series of healthy subjects interbeat electrocardiogram of $6000$ data \cite{gold2}. These are
a prototypical example of a long-range correlated stochastic process \cite{gold1}. The straight line characterizes the theoretical result for an uncorrelated process. The degree distribution is exponential with $\lambda=0.5>\lambda_{un}$,
corresponding to a correlated stochastic process, as predicted by our theory. Results correspond to an average over five time series, one of them being depicted in the left panel.\label{fig2b}} 
\end{figure}

As a further example, we use the dynamics of healthy sinus rhythm cardiac interbeats, a physiological stochastic process that has been shown to evidence long-range
correlations \cite{gold1}. In figure \ref{fig2b} we have plotted the degree distribution of the HVG generated by a time series of
the beat-to-beat fluctuations of five young subjects (21-34 yr) with healthy sinus rhythm heartbeat \cite{gold2}. Even if these time series are short (about $6000$ data), 
the results match those obtained in the previous examples, namely, that the associated graph is characterized by an exponential degree distribution with slope $\lambda>\lambda_{un}$,
as it corresponds to a correlated stochastic process.

All these examples provide evidence showing that a time series of stochastic correlated data can be characterized by its associated HVG. This graph has an exponential
node-degree distribution with a characteristic parameter $\lambda$ that always exceeds the uncorrelated value $\lambda_{un}=\ln(2/3)$. This is true even in the case of 
weakly correlated processes (large values of the correlation exponent $\gamma$ in the case of power-law, long-range, decay of correlations, or small values of  $\tau$ in the case of an exponential, short-range, decay).

\section{Chaotic maps}

\begin{figure}[]
\centering
\includegraphics[width=1.0\textwidth]{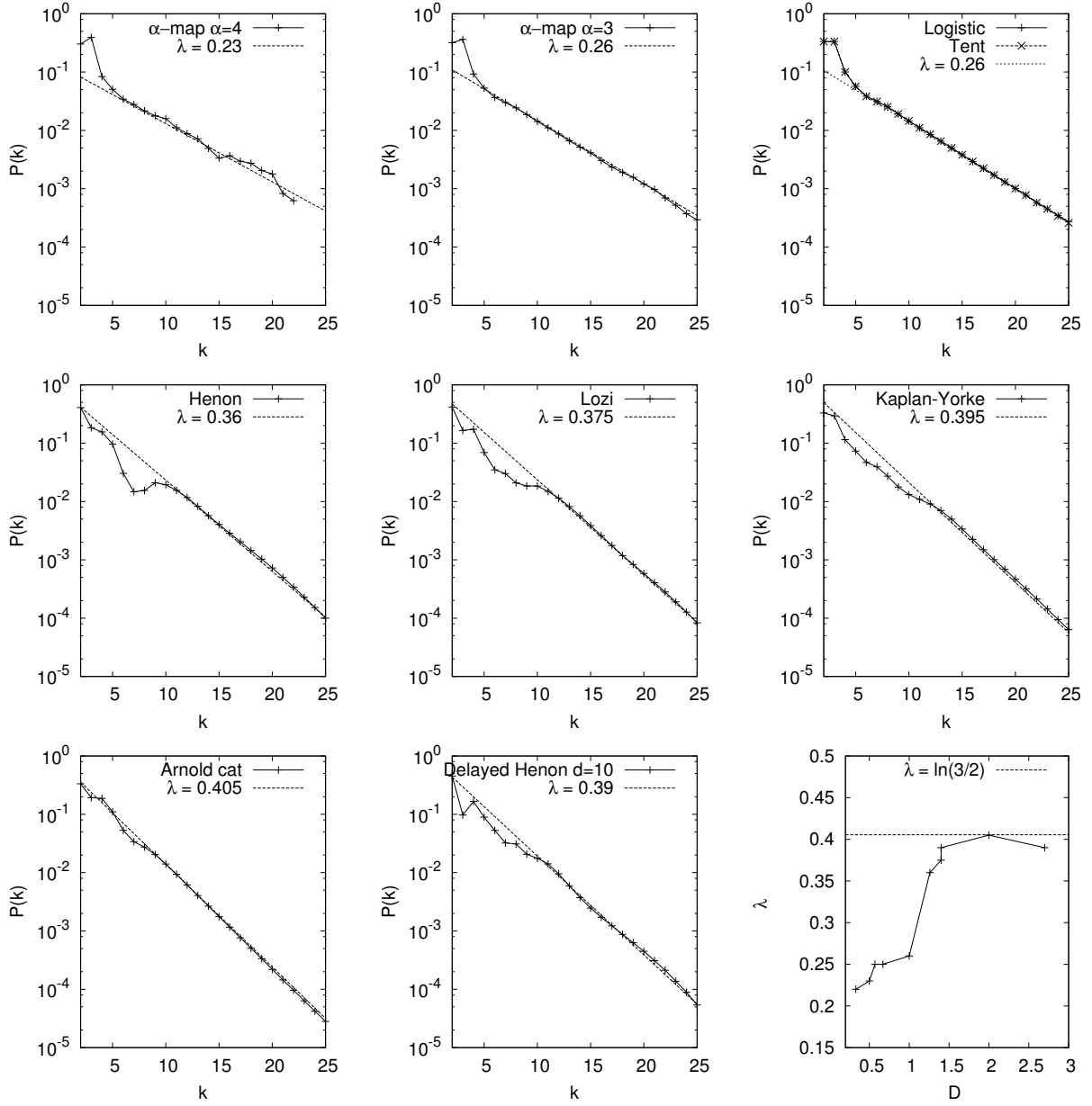}
\caption{From left to right, up to bottom: Semilog plot of the degree distributions of Horizontal visibility graphs associated to series
generated through chaotic maps with different correlation dimension (data are averaged over $100$ realizations). In every case we find that the degree distribution is exponential $P(k)\sim \exp(-\lambda k)$, where
the slope $\lambda$ monotonically increases with the correlation dimension $D$. In the bottom right we plot the functional relation between $\lambda$ and $D$, showing
that the values of $\lambda$ converge towards the uncorrelated situation ($\lambda=\lambda_{un}=\ln(3/2)$) for increasing values of the chaos dimensionality.} \label{multi_ch}
\end{figure}

 We now focus on processes generated by chaotic maps. In a preceding work \cite{PRE}, we conjectured that the Poincar\'{e} recurrence theorem
suggests that the degree distribution of HVGs associated to chaotic series should be asymptotically exponential. Here we address several deterministic
time series generated by chaotic maps, and analyze the possible deviations from the uncorrelated results. Concretely, we tackle the following maps:\\

\noindent(1) the $\alpha$-map $f(x)=1-|2x-1|^{\alpha}$, that reduces to the logistic and tent maps in their fully chaotic region for $\alpha=2$ and $\alpha=1$ 
respectively, for different values of $\alpha$,\\
(2) the 2D H\'{e}non map ($x_{t+1}=y_t+1-ax_t^2$, $y_{t+1}=bx_t$) in the fully chaotic region ($a=1.4$, $b=0.3$);\\
(3) a time-delayed variant of the H\'{e}non map: $x_{t+1}=bx_{t-d}+1-ax_t^2 $ in the region ($a=1.6$, $b=0.1$), where it shows
chaotic behavior with an attractor dimension that increases linearly with the delay $d$ \cite{EJTP}. This model has also been used
for chaos control purposes \cite{chaos-control}, although here we set the parameters $a$ and $b$ to values for which we find high-dimensional chaos
for almost every initial condition \cite{EJTP};\\
(4) the Lozi map, a piecewise-linear variant of the H\'{e}non map given by $x_{t+1}=1+y_n-a|x_t|,\  y_{t+1}=bx_t$ in the chaotic regime $a=1.7$ and $b=0.5$;\\
(5) the Kaplan-Yorke map $x_{t+1}=2x_t \mod(1), y_{t+1}=\lambda y_t + \cos(4\pi x_t) \mod(1)$; and\\
(6) the Arnold cat map $x_{t+1}=x_t+y_t \mod(1), y_{t+1}=x_t+2y_t \mod(1)$, a conservative system with integer Kaplan-Yorke dimension. References for these maps
can be found in \cite{sprott2}. 

\begin{figure}[h]
\centering
\includegraphics[width=0.65\textwidth]{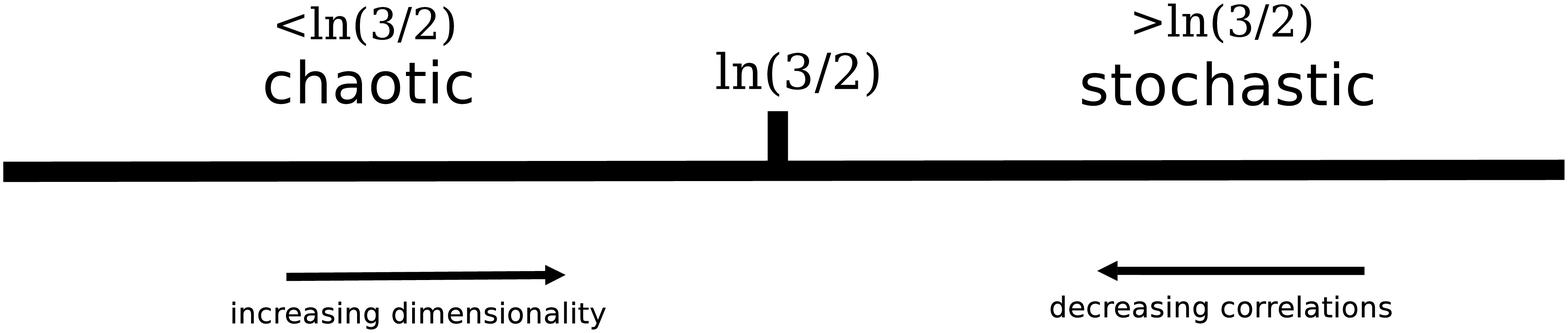}
\includegraphics[width=0.5\textwidth]{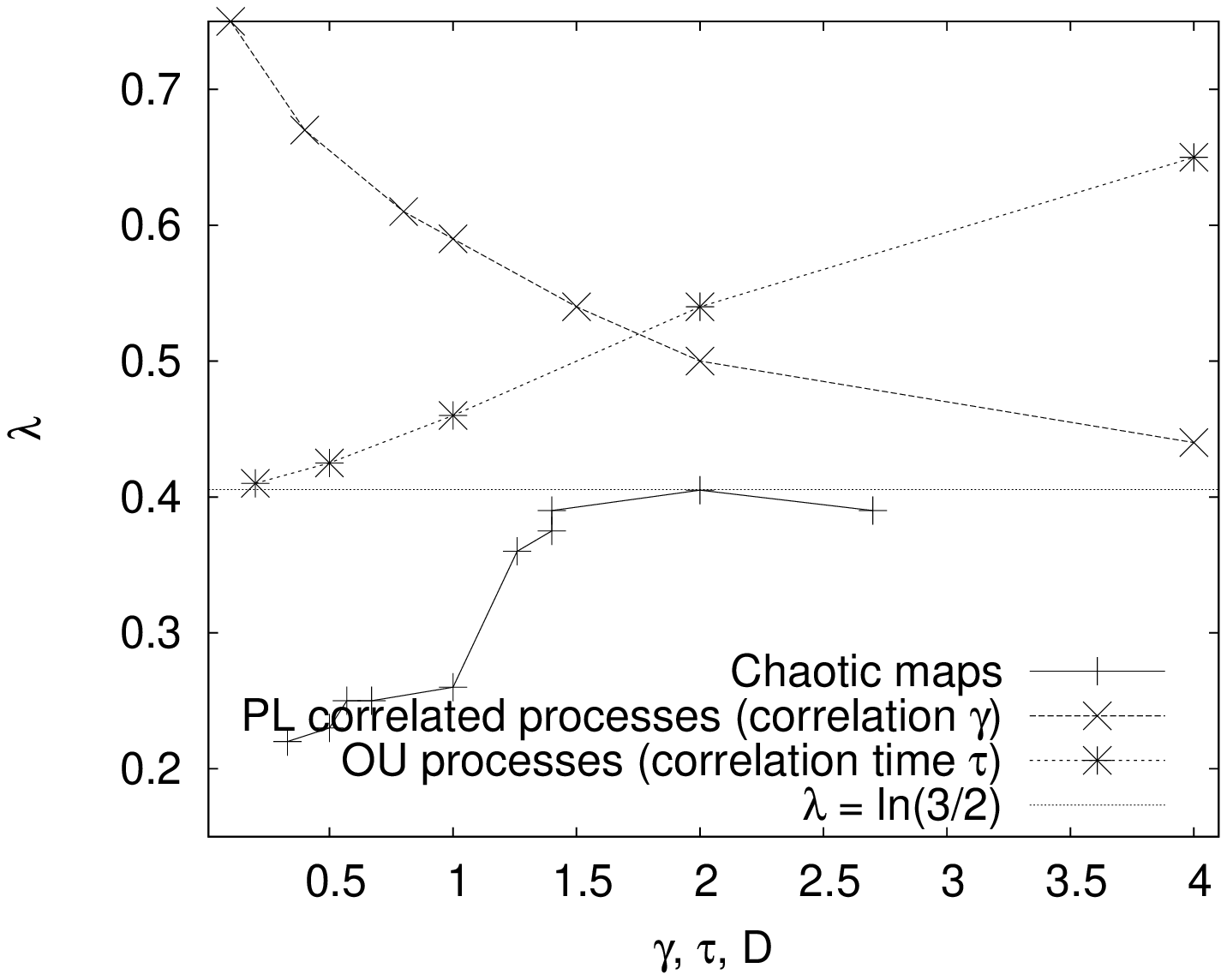}
\caption{(\textit{Left}) $\lambda$ diagram: for $\lambda<\ln(3/2)$, we have a chaotic process, whereas $\lambda>\ln(3/2)$ corresponds to a correlated stochastic process. The frontier value $\lambda=\ln(3/2)$ corresponds to the uncorrelated case. Note that this latter value is an exact result of the theory \cite{PRE}. (\textit{Right}) Plot of the values of $\lambda$ for several processes, namely: (i) for power-law correlated stochastic series with correlation function $C(t)=t^{-\gamma}$, as a function of the
correlation $\gamma$, (ii) for Ornstein-Uhlenbeck series with correlation function $C(t)=\exp(-t/\tau)$, as a function of the correlation time $\tau$, and (iii) for
different chaotic maps, as a function of their correlation dimension $D$. Errors in the estimation of $\lambda$ are incorporated in the size of the dots. Notice that stochastic processes cluster in the region $\lambda>\lambda_{un}$ whereas
chaotic series belong to the opposite region $\lambda<\lambda_{un}$, evidencing a convergence towards the uncorrelated value $\lambda_{un}=\ln(3/2)$ \cite{PRE} for
decreasing correlations or increasing chaos dimensionality respectively.} \label{diagram}
\end{figure}

In figure \ref{multi_ch} we plot in semi-log the degree distribution of chaotic series of $2^{18}$ data generated through several chaotic maps (logistic, tent, $\alpha$-map
with $\alpha=3$ and $4$, H\'{e}non, delayed H\'{e}non with a delay $d=10$, Lozi, Kaplan-Yorke and Arnold cat). We find that the tails of the degree distribution can be well
approximated by an exponential function $P(k)\sim \exp(-\lambda k)$. Remarkably, we find that $\lambda<\lambda_{un}$ in every case, where $\lambda$ seems to increase monotonically
as a function of the chaos dimensionality \cite{comment}, with an asymptotic value $\lambda\rightarrow \ln(3/2)$ for large values of the attractor dimension (see the
right-hand side bottom of the figure where we plot the specific values of $\lambda$ as a function of the correlation dimension of the map \cite{sprott2}). 
Again, we deduce that the degree distribution for uncorrelated series is
a limiting case of the degree distribution for chaotic series but, as opposed to what we found for stochastic processes, the convergence flow towards $\lambda_{un}$ is from below, 
and therefore $\lambda=\ln(3/2)$ plays the role of an effective frontier between correlated stochastic and chaotic processes (see left part of Fig. \ref{diagram} for an illustration).

A summary of all data series analyzed can be seen in the right panel of Fig. \ref{diagram}, where we plot the fitted slope $\lambda$ of particular series generated through 
power-law correlated (as a function of correlation $\gamma$) and exponentially correlated (as a function of correlation time $\tau$) stochastic processes, and through the aforementioned chaotic
maps (as a function of the correlation dimension $D$). In the following sections we will provide some analytical developments and heuristic arguments supporting our findings.

\section{Heuristics}
 We argue first that correlated series show lower data variability than uncorrelated ones, so decreasing the possibility of a node to
reach far visibility and hence decreasing (statistically speaking) the probability of appearance of a large degree. Hence, the correlation tends to decrease
the number of nodes with large degree as compared to the uncorrelated counterpart. Indeed, in the limit of infinitely
large correlations ($\gamma\rightarrow0$ or $\tau\rightarrow\infty$), the variability reduces to zero and the series become constant. The degree distribution in this
limit case is, trivially, $$P(k)=\delta(k-2)=\lim_{\lambda\rightarrow\infty}\frac{\lambda}{2}\exp(-\lambda|k-2|),$$ that is to say, infinitely large correlations would be associated
to a diverging value of $\lambda$. This tendency is on agreement with the numerical simulations (right panel of figure \ref{diagram}) where we show that $\lambda$
monotonically increases with decreasing values of $\gamma$ or increasing values of $\tau$ respectively. Having in mind that in the limit of small correlations the
theorem previously stated implies that $\lambda\rightarrow\lambda_{un}=\ln(3/2)$, we can therefore conclude that for a correlated stochastic process $\lambda_{stoch}>\lambda_{un}$.

Concerning chaotic series, remember that they are generated through a
deterministic process whose orbit is continuous along the attractor. This continuity introduces a smoothing effect in the series that, statistically speaking,
increases the probability of a given node to have a larger degree (uncorrelated series are rougher and hence it is more likely to have more nodes with smaller degree).
Now, since in every case we have exponential degree distributions (this fact being related with the Poincar\'{e} recurrence theorem for chaotic series and with the
return distribution in Poisson processes for stochastic series \cite{PRE}), we conclude that the deviations must be encoded in
the slope $\lambda$ of the exponentials, such that $\lambda_{chaos}<\lambda_{un}<\lambda_{stoch}$, in good agreement with our numerical results.

\section{Analytical developments}
In \cite{PRE} we proved that $P(k)=(1/3)(2/3)^{k-2}$ for uncorrelated random series. To find out a similar closed expression in the case of generic
chaotic or stochastic correlated processes is a very difficult task, concretely since variables can be long-range correlated and hence the probabilities
cannot be separated
(lack of independence). This leads to a very involved calculation which is typically impossible to solve in the general case. However, some analytical developments
can be made in order to compare them with our numerical results. Concretely, for Markovian systems global dependence is reduced to a one-step dependence. We will
make use of such property to derive exact expressions for $P(2)$ and $P(3)$ in some Markovian systems (both deterministic and stochastic).
In order to compare the theoretical calculations of $P(2)$ and $P(3)$ in the case of an Ornstein-Uhlenbeck process
(detailed in section III) with the numerical
results, in table \ref{table1} we have depicted the associated numerical results for different correlation times. 

\begin{table}
\begin{ruledtabular}
\begin{tabular}{cccccddd}
$\tau$&$P_{OU}(2)$&$P_{OU}(3)$&$P_{log}(2)$&$P_{log}(3)$\\
\hline
1.0&0.3012&0.232&-&-\\
0.5&0.3211&0.227&-&-\\
0.1&0.3333&0.222&-&-\\
-&-&-&0.3333&0.3332\\
\end{tabular}
\end{ruledtabular}
\caption{\label{table1}Numerical results of $P(2)$ and $P(3)$ associated to (i) an Ornstein-Uhlenbeck series of $N=2^{18}$ data with correlation function $C(t)=\exp(-t/\tau)$,
for different values of the correlation time $\tau$, and (ii) to a series of $N=2^{18}$ data extracted from a logistic map in its fully chaotic region, $\alpha$-map with $\alpha=2$.
To be compared with exact results derived in section VI.}
\end{table}

\subsection{Ornstein-Uhlenbeck process} 
Suppose a short-range correlated series (exponentially decaying correlations) of infinite size generated through an Ornstein-Uhlenbeck process, and generate its 
associated HVG. Let us consider the probability that a node chosen at random has degree $k=2$. This node is associated to a datum labelled $x_0$ without lack of
generality.
Now, this node will have degree $k=2$ if the datum
first neighbors, $x_1$ and $x_{-1}$ have values larger than $x_0$:
$$P(k=2)=P(x_{-1}>x_0 \cap x_1>x_0)$$
If series data were random and uncorrelated, we would have
\begin{equation}
 P_{un}(2)=\int_{-\infty}^\infty dx_{0}\, f(x_{0}) \int_{x_0}^\infty dx_{-1}\, f(x_{-1})\int_{x_0}^\infty dx_{1}\,f(x_{1})=1/3,
\end{equation}
where we have used the properties of the cumulative probability distribution (note that this result holds for any continuous probability density $f(x)$, as shown
in \cite{PRE}). Now, in our case the variables are correlated, so in general we should have
\begin{equation}
 P_{OU}(2)=\int_{-\infty}^\infty dx_{0}\, \int_{x_0}^\infty dx_{-1}\,\int_{x_0}^\infty dx_{1}\, f(x_{-1},x_0,x_1).\label{P2generico}
\end{equation}
We use  the Markov property  $f(x_{-1},x_0,x_1)=f(x_{-1})f(x_{0}|x_{-1}) f(x_{1}|x_0)$, that holds for
an Ornstein-Uhlenbeck process with correlation function $C(t)\sim\exp(-t/\tau)$\cite{kampen}:
\begin{equation}
 f(x)=\frac{\exp(-x^2/2)}{\sqrt{2\pi}} ,\hspace{1.0cm}  f(x_2|x_1)= \frac{\exp(-(x_2-Kx_1)^2/2(1-K^2))}{\sqrt{2\pi(1-K^2)}},\label{f2}
\end{equation}
where $K=\exp(-1/\tau)$.\\

 Numerical integration allows us to calculate $P_{OU}(2)$ for every given value of the correlation time $\tau$. For instance,
we find $P_{OU}(2)|_{\tau=1.0}=0.3012$, $P_{OU}(2)|_{\tau=0.5}=0.3211$, $P_{OU}(2)|_{\tau=0.1}=0.3331$, in perfect agreement
with our previous numerical results (see table \ref{table1}).
\begin{figure}[h]
\centering
\includegraphics[width=0.8\textwidth]{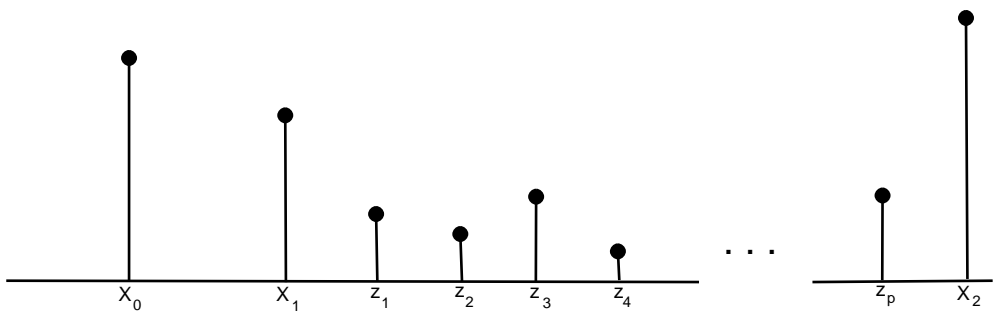}
\caption{Schematic representation of a situation where datum $x_0$ has right-visibility of two data ($P_+(2)$), $x_1$ and $x_2$. An arbitrary number of hidden data
can be placed between $x_1$ and $x_2$, and this has to be taken into account in the calculation of $P(3)$.} \label{P2fig}
\end{figure}

 An arbitrary datum $x_0$ of a series extracted from an Ornstein-Uhlenbeck will have an associated node with degree $k=3$ with a certain probability
$P_{OU}(3)$ which is the sum of the probabilities associated to two possible scenarios, namely (i) the probability that $x_0$ has two visible data in its right-hand side and a
single one in its left-hand side, labeled $P_{OU}^+(3)$, and (ii) the probability that $x_0$ has two visible data in its left-hand side and a single one in
its right-hand side, labeled $P_{OU}^-(3)$. In the very particular case of stationary Markovian processes (such as the Ornstein-Uhlenbeck), time invariance yields
$P_{OU}(3)=2P_{OU}^+(3)$. Let us tackle now the calculation of $P_{OU}^+(3)$. Let us quote $x_{1},x_{2}$ the right-hand side visible data of
$x_0$ and $x_{-1}$ the left-hand side visible one. Formally, we have 
\begin{equation}
P_{OU}^+(3)= \int_{-\infty}^\infty dx_0\, \int_{x_0}^\infty dx_{-1}\, f(x_{-1})f(x_0|x_{-1})P_+(2|x_0),\label{P3}
\end{equation}
where $P_+(2|x_0)$ is the probability that $x_0$ sees two data on its right-hand side (see figure \ref{P2fig} for a graphical illustration). Of course in $P_+(2|x_0)$ we have to take
into account the possibility of having an arbitrary number of hidden (non visible) data between the first and the second visible datum, so
\begin{eqnarray}
P_+(2|x_0)&=&\int_{-\infty}^{x_0}dx_1\int_{x_0}^\infty dx_2f(x_1|x_0)f(x_2|x_1) +\\
&&\int_{-\infty}^{x_0}dx_1\int_{-\infty}^{x_1} dz_1 \int_{x_0}^{\infty} dx_2 f(x_1|x_0)f(z_1|x_1)f(x_2|z_1) + \nonumber\\
&&\int_{-\infty}^{x_0}dx_1\int_{-\infty}^{x_1} dz_1\int_{-\infty}^{x_1} dz_2  \int_{x_0}^{\infty} dx_2 f(x_1|x_0)f(z_1|x_1)f(z_2|z_1)f(x_2|z_2)+...\nonumber\\
&\equiv&\sum_{p=0}^\infty I(p|x_0)\nonumber
\label{P+2OU}
\end{eqnarray}
where $f(x|y)$ is the Ornstein-Uhlenbeck transition probability defined in equation \ref{f2}, and $z_p$ is the $p$-th hidden data located between $x_1$ and $x_2$
(note that there can be an eventually infinite amount of hidden data between $x_1$ and $x_2$ and these configurations have to be taken into account in the calculation).
Here $I(p|x_0)$ characterizes the probability that $x_0$ sees two data on its right-hand side with $p$ hidden data between them. 

 A little algebra allows us to write 
\begin{equation}
I(p|x_0)=\int_{-\infty}^{x_0} dx_1f(x_1|x_0) G_p(x_1,x_1,x_0),\label{generalI} 
\end{equation}
where the function $G_p$ satisfies a recursive relation:
\begin{eqnarray}
G_0(x,y,z)&\equiv&\int_z^\infty f(h|y)dh, \\
G_p(x,y,z)&=&\int_{-\infty}^x dh f(h|y)G_{p-1}(x,h,z), \ p\geq1. \label{Ggen}
\end{eqnarray}
This is a convolution-like equation that can be formally rewritten as $G_p=TG_{p-1}$, or $G_p=T^pG_0$,
with an integral operator $T=\int_{-\infty}^x dh f(h|y)$. Accordingly, we have

\begin{equation}
P_+(2|x_0)=\int_{-\infty}^{x_0} dx_1f(x_1|x_0) \sum_{p=0}^{\infty}G_p(x_1,x_1,x_0)\equiv\int_{-\infty}^{x_0} dx_1f(x_1|x_0) S(x_1,x_1,x_0),
\label{p2m}
\end{equation}
 where we have defined the summation $S(x,y,z)$ as
\begin{equation}
 S(x,y,z)=\sum_{p=0}^\infty G_p(x,y,z)=\sum_{p=0}^\infty T^p G_0=\frac{1}{1-T}G_0,\label{S1}
\end{equation}
where in the last equality we have used the summation and convergence properties of geometric series (Picard sequence).
This is valid whenever the spectral radius of the linear operator $r(T)<1$, that is, if
\begin{equation}
\lim_{n\rightarrow\infty} \bigg[||T^n||\bigg]^{1/n}<1,
\end{equation}
where $||T||=\max_{y\in(-\infty,x)} \int_{-\infty}^x dh |f(h|y)|$ is the norm of $T$. Now, this condition is trivially fulfilled given the fact that $f(x|y)$ is a
Markov transition probability. Then equation \ref{S1} can be written as $(1-T)S=G_0$, or more concretely
\begin{equation}
S(x,y,z)={G}_0(x,y,z) + \int_{-\infty}^x dh f(h|y) S(x,h,z), \label{SSS}
\end{equation}
which is a Volterra equation of the second kind \cite{librodeernesto} for $S(x,y,z)$. Note that it can also be seen as a multidimensional convolution-like equation since
the argument in the Markov transition probability $f(h|y)$ has the shape $h-y'$, where $y'=\exp(-1/\tau)y$. Hence $f$ can be understood as the kernel of the convolution.

Typical one-dimensional Volterra integral equations can be numerically solved applying quadrature formulae
for approximate the integral operator \cite{librodeernesto}. The technique can be easily extended whenever the integral equation involves more than one variable, as it is our case.
Specifically, a Simpson-type integration scheme leads to a recursion relation with a step $\delta$ to compute the function $S(x,y,z)$. One technical point is that one needs to replace the $-\infty$ limit in the integral by a sufficienly small number $a$. We have found that $a=-10$ is enough for a good convergence of the algorithm. Given a value of $z$ the recursion relation  
\begin{eqnarray}
&&S(a,a+n\delta,z)= G_0(a,a+n\delta,z)\nonumber \\
&&S(a+k\delta,a+n\delta,z)= G_0(a,a+n\delta,z)+\delta\sum_{i=0}^{k-1}f(a+i\delta|a+n\delta)S(a+(k-1)\delta,a+i\delta,z)+ O(\delta^2),\label{SS}
\end{eqnarray}
for $k=0,1,2,\dots$ and $n=0,1,\dots,k$, allows us to compute $S(x,y,z)$ for $y\le x$.

Summing up, the procedure to compute $P_{OU}(3)$ is the following: calculate $S(x_1,x_1,x_0)$ using the previous recursion relation and use this value to obtain $P_+(2|x_0)$ from a numerical integration of the right-hand-side of equation \ref{p2m}) (again, the lower limit will be replaced by $x_1=a$ ). Finally, integrate numerically equation \ref{P3} to obtain $P^+_{OU}(3)$ from which it readily follows $P_{OU}(3)$. 
Applying this methodology for an integration step $\delta=4\times 10^{-3}$ we find $P_{OU}(3)|_{\tau=1.0}=0.230$, $P_{OU}(3)|_{\tau=0.5}=0.226$, or $P_{OU}(3)|_{\tau=0.1}=0.221$,
in good agreement with numerical results (table \ref{table1}).

\subsection{Logistic map} 
A chaotic map of the form $x_{n+1}=F(x_n)$ does also have the Markov property, and therefore a similar analysis can therefore apply (even if chaotic maps are deterministic). For chaotic dynamical systems whose trajectories belong to the attractor, there exists a probability measure that characterizes the long-run proportion of time spent by the system in the various
regions of the attractor. In the case of the logistic map $F(x_n)=\mu x_n(1-x_n)$ with parameter  $\mu = 4$, the attractor is the whole
interval $[0,1]$ and the probability measure $f(x)$ corresponds to the \textit{beta} distribution with parameters  $a = 0.5$  and  $b = 0.5$:
\begin{equation}
f(x)=\frac{x^{-0.5}(1-x)^{-0.5}}{\texttt{B}(0.5,0.5)}. 
\label{rho}
\end{equation}
Now, for a deterministic system, the transition probability is
\begin{equation}
f(x_{n+1}|x_n)=\delta(x_{n+1}-F(x_n)), \label{rho2}
\end{equation}
where $\delta(x)$ is the Dirac delta distribution. Departing from equation \ref{P2generico}, for the logistic map $F(x_n)=4x_n(1-x_n)$ and $x_n \in [0,1]$, we have
\begin{eqnarray}
P_{log}(2)&=&\int_{0}^1 dx_{0}\int_{x_0}^1 f(x_{-1})f(x_{0}|x_{-1}) dx_{-1}\int_{x_0}^1 f(x_{1}|x_0)dx_{1}=\nonumber \\
&&\int_{0}^1 dx_{0}\int_{x_0}^1 f(x_{-1})\delta(x_0-F(x_{-1}))dx_{-1}\int_{x_0}^1 \delta(x_1-F(x_0))dx_{1}.\label{log2}
\end{eqnarray}
Now, notice that, using the properties of the Dirac delta distribution, 
$\int_{x_0}^1 \delta(x_1-F(x_0))dx_{1}$ is equal to one iff $F(x_0)\in[x_0,1]$, what will happen iff $0<x_0<3/4$, and zero otherwise. Therefore the only effect of
this integral is to restrict the integration range of $x_0$ to be $[0,3/4]$.

On the other hand,
$$\int_{x_0}^1 f(x_{-1})\delta(x_0-F(x_{-1}))dx_{-1}=\sum_{x^*_k|F(x^*_k)=x_0}f(x^*_k)/|F'(x^*_k)|,$$
that is, the sum over the roots of the equation $F(x)=x_0$, iff $F(x_{-1})>x_0$. But since $x_{-1}\in[x_0,1]$ in the latter integral,
it is easy to see that again, this is verified iff $0<x_0<3/4$ (as a matter of fact, if $0<x_0<3/4$ there is always a \textit{single} value of $x_{-1}\in[x_0,1]$ such that $F(x_{-1})=x_0$, so the sum restricts to the adequate root). It is easy to see that the particular value is $x^*=(1+\sqrt{1-x_0})/2$.
Making use of these piecewise solutions and equation \ref{rho}, we finally have
\begin{equation}
P_{log}(2)=\int_0^{3/4} \frac{f(x^*)}{4\sqrt{1-x_0}}dx_0=1/3,
\end{equation}
which is in perfect agreement with the numerical results (see table \ref{table1}). Note that a similar development can be fruitfully applied to other chaotic maps, provided
that they have a well defined natural measure.

 The approach for analytically calculating $P_{log}^+(3)$ in the case of a chaotic map with a well defined natural measure --such as the logistic map in its fully
chaotic region $\mu=4.0$-- is very similar to the one adopted for an Ornstein-Uhlenbeck process, again replacing the probability density and Markovian
transition probability with equations \ref{rho} and \ref{rho2}. Remarkably, applying the properties of the Dirac delta and the logistic map
it can be easily proved that $I(0)=1$ and $I(p)=0\ \forall p>0$ provided that $x_0$ is restricted to the
range $3/4<x_0<1$. The whole calculation therefore reduces to
\begin{equation}
P_{log}^+(3)=\int_{3/4}^1 \frac{f(x^*)}{4\sqrt{1-x_0}}dx_0=1/6, 
\end{equation}
that yields $P_{log}(3)=2P_{log}^+(3)=1/3$, in perfect agreement with numerical results (see table \ref{table1}). Again, similar developments can be straightforwardly
applied to other chaotic maps with well defined natural measure.

\section{Comment on noisy periodic maps}
\begin{figure}[h]
\centering
\includegraphics[width=1.0\textwidth]{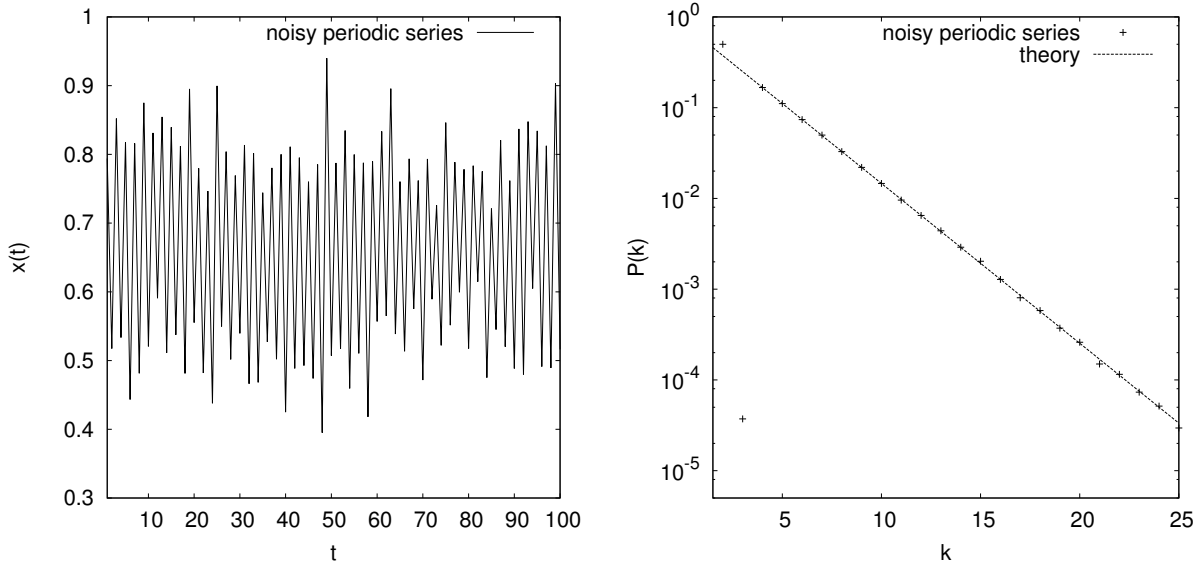}
\caption{\textit{Left}: Periodic series of $2^{20}$ data generated through the logistic map $x_{n+1}=\mu x_n(1-x_n)$ for $\mu=3.2$ (where the map shows periodic behavior with period $2$)
polluted with extrinsic white gaussian noise extracted from a Gaussian distribution $N(0,0.05)$. \textit{Right}: Dots represent the degree distribution of the associated HVG, whereas
the straight line is equation \ref{teoriaperi} (the plot is in semi-log). Note that $P(2)=1/2$, also as theory predicts, and that $P(3)$ is not exactly zero due to boundary effects in the time series.
The algorithm efficiently detects both signals and therefore easily distinguishes extrinsic noise.} \label{ejemplo}
\end{figure}
Periodic series have an associated HVG with a degree distribution formed by a finite number of peaks,
these peaks being related to the series period, what is reminiscent of the discrete Fourier spectrum of a periodic series \cite{PNAS, PRE}.
The reason is straightforward: a periodic series maps into an HVG
which, by construction, is a repetition of a root motif. Now, if we superpose a small amount of noise to a periodic series (a so-called \textit{extrinsic} noise),
while the degree of the nodes with associated small values will remain rather similar, the nodes associated to higher values will eventually increase their visibility and hence
reach larger degrees. Accordingly, the delta-like structure of the degree distribution will be perturbed, and an exponential
tail will arise due to the presence of such noise. Can the algorithm characterize such kind of series? The answer is positive, since the degree distribution can
be analytically calculated as it follows:
Consider for simplicity a period-2 time series polluted with white noise (see the left part of figure \ref{ejemplo} for a graphical illustration).
The HVG is formed by two kind of nodes: those associated to high data with values ($(i_1,i_3,i_5,...)$ in the figure) and those associated to data with small values
($(i_2,i_4,i_6,...)$). These latter nodes will have, by construction, degree $k=2$. On the other hand, the subgraph formed by the odd nodes $(i_1,i_3,i_5,...)$ will essentially reduce to the one 
associated to an uncorrelated series, \textit{i.e.} its degree distribution will follow equation \ref{pk}. Now, considering the whole graph, the resulting degree distribution will
be such that
\begin{eqnarray}
&&P(2)=1/2,\nonumber\\
&&P(3)=0,\nonumber\\
&&P(k+2)=\frac{1}{3}\bigg(\frac{2}{3}\bigg)^{k-2}, \ k\geq2, \nonumber\\
&&\Leftrightarrow P(k)= \frac{1}{4}\bigg(\frac{2}{3}\bigg)^{k-3}, \ k\geq4,\label{teoriaperi}
\end{eqnarray}
that is to say, introducing a small amount of extrinsic uncorrelated noise in a periodic signal introduces an exponential tail in the HVG's degree distribution with slope $\ln(3/2)$.
In the left part of figure \ref{ejemplo} we plot in semi-log the degree distribution of a periodic-2 series of $2^{20}$ data polluted with an extrinsic white Gaussian noise
extracted from a Gaussian distribution $N(0,0.05)$. Numerical results confirm the validity of equation \ref{teoriaperi}. Note that
this methodology can be extended to every integrable deterministic system, and therefore we conclude that extrinsic noise in a mixed time series is well captured by the algorithm.\\
Conversely, introducing a small amount of
\textit{intrinsic} noise in a periodic series is more tricky. For instance, consider the noisy logistic map defined as
$$x_{t+1}=\mu x_t(1-x_t) + \sigma \xi_t,$$
where $\xi_t$ are independent random numbers extracted from a Gaussian distribution $N(0,\sigma)$ with zero mean and standard deviation$\sigma$. 
For some values of $\mu<\mu_{\infty}$ (that is, in the periodic regime of the associated noise-free logistic map), small amounts of intrinsic noise can produce orbits very similar to those generated by
the noise-free version
of the map in the chaotic regime \cite{vulpiani}, in the sense that, superposed to the delta-like shape of $P(k)$, an asymptotic exponential
tail with $\lambda<\lambda_{un}$ may eventually develop. Besides the delta-like structure of $P(k)$ appearing for short values of $k$ (reminiscent of the periodicity of the noise-free
map), the algorithm fails in determining the source of entropy of the system (which is stochastic here, and therefore $\lambda\geq\lambda_{un}$). This is a typical pathological case \cite{vulpiani, cencini}
where chaos and noise are difficult to distinguish. Indeed, it has been pointed out that rather sophisticated methods such as finite size Lyapunov exponents or $(\epsilon,\tau)-$entropies have difficulties to
determine the chaotic/stochastic nature of these maps for finite resolution \cite{cencini}. This limitation of the algorithm should be investigated in detail in further work. 

\section{Conclusion}
To conclude, we have shown that correlated stochastic series
map into an horizontal visibility graph with an exponential degree distribution with slope $\lambda>\ln(3/2)$, that slowly tends to its asymptotic value for very weak correlations.
Results are confirmed for a real physiological time series that has been previously shown to evidence long-range correlations \cite{gold1}. Similar
results have been obtained for the case of chaotic series, with the peculiarity that the slope of the degree distribution converges to $\ln(3/2)$ in the opposite direction ($\lambda<\ln(3/2)$).
In a preceding work we analytically proved that for an uncorrelated random series, the slope is exactly $\ln(3/2)$, independently of
the probability density. We therefore conclude that chaotic maps and correlated stochastic processes seem to belong to different regions of the $\lambda$ diagram, where
$\lambda=\ln(3/2)$ plays the role of an effective frontier between both processes.
It is worth commenting that the horizontal visibility algorithm is very fast
(as a guide, the generation of the associated graph  for a series of $N=2^{18}$ data in a standard personal computer
takes a computation time of the order of a few seconds). Applications include direct characterization of complex signals
such as physiological series or series extracted from natural phenomena, as a first step where to discriminate amongst several modeling framework approaches. Questions for future work include
a deeper characterization of this method, concretely the incorporation of Lyapunov exponents and the associated short-term memory effects within the visibility framework,
and the study of noisy maps, which hitherto constitute a  limitation of the algorithm. 
The characterization of flows that
produce continuous time series is also an open problem for future research.

\textbf{Acknowledgments}
We thank Bartolo Luque for fruitful discussions and acknowledge financial support by the MEC (Spain) and FEDER (EU) through projects FIS2007-60327 and FIS2009-13690.

\section{Appendix: Statistical error in $\lambda$}
The calculation of $\lambda$ comes straightforwardly from the fitting of the HVG's degree distribution (concretely, the tail) to an exponential function. Two possible sources of uncertainty in this calculation are present, namely: (i) the finite size effects associated to finite time series induce a lack of statistics for large values of the graph degree $k$, and (ii) experimental time series are often polluted with measurement errors. While a detailed and systematic analysis of these issues is beyond the scope of this work, at this point we can outline the following comments: (i) for the stochastic correlated and chaotic systems considered in this work, finite size effects seem to be unrelevant for relatively large time series ($N>2^{14}$), in the sense that the region $[k_{min}, k_{max}]$ where an exponential function is very well fitted is large enough (see figures \ref{multi_PL} and \ref{multi_ch}) and accordingly, the error associated to $\lambda$ can be simply estimated as the error of the exponential fitting. (ii) In general, the procedure to check the effect of finite size is the following: consider a stationary series of $N$ data. The error in the calculation of $\lambda$ can be estimated by partitioning the original series in $s$ samples of $N/s$ data, labelled $s_1,...,s_{N/s}$. Accordingly, each series generates an HVG whose degree distribution can be fitted to an exponential function with slope $\lambda_{s_i}$, such that $<\lambda>=\frac{1}{s}\sum_{i=1}^s\lambda_{s_i}$ and the associated error is simply the standard deviation from the mean (this is equivalent to performing a time average). (iii) In practice, this latter procedure is not appropiate for very short time series ($N$ of order $O(10^3)$); in this case an ensemble average is better suited (note at this point that stationarity is needed in order to guarantee that averaging over ensembles and over time yield equivalent results).

\begin{figure}[h]
\centering
\includegraphics[width=0.8\textwidth]{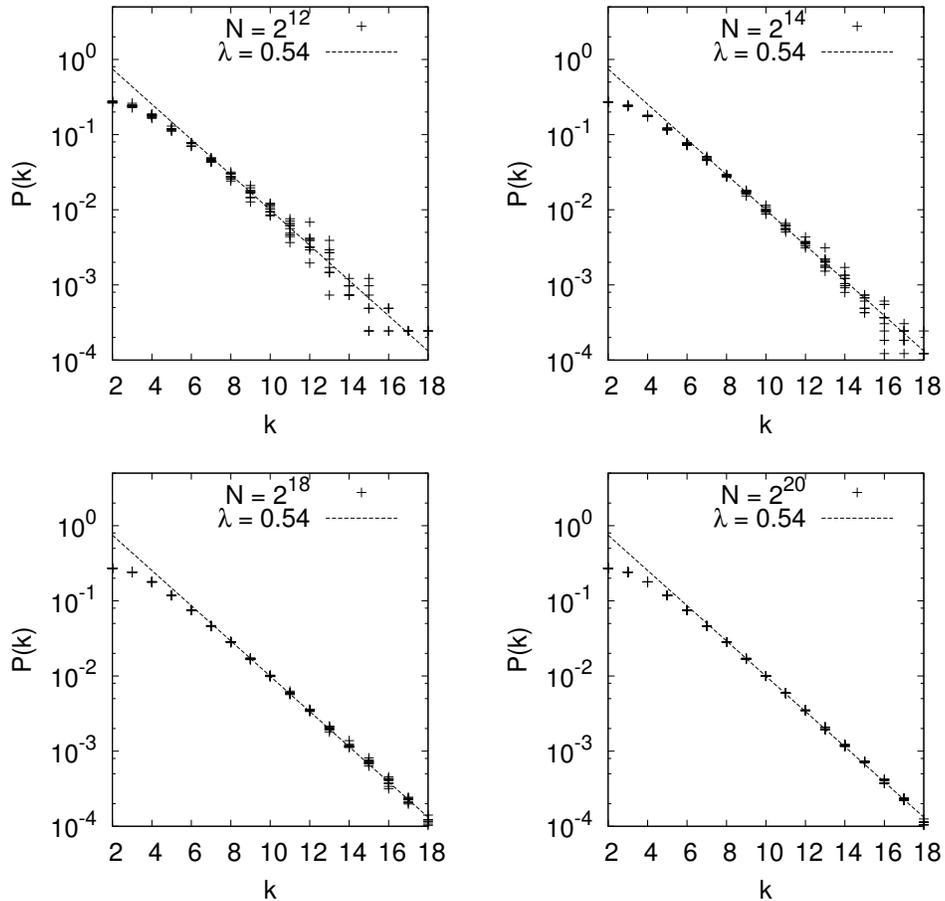}
\caption{Semi-log plots of the degree distributions of power-law correlated stochastic series of different sizes with correlation $\gamma=1.5$ (see figure \ref{multi_PL}). Crosses represent the degree distribution of the associated HVG for each realization of the stochastic process ($10$ time series). Notice that statistical deviations from $\lambda=0.54$ (which are more acute in the tail of the distribution, where statistics are poor) decrease with system's size.} \label{error}
\end{figure}

For illustration purposes, we address the case of a power-law correlated stochastic process with correlation function $C(t)=t^{-\gamma}$, with $\gamma=1.5$, for which a previous analysis shows that the associated HVG has an exponential degree distribution with slope $\lambda=0.54$ (see figure \ref{multi_PL}). For a given time series size $N$, we generate ten series and plot the degree distribution of the associated HVGs in figure \ref{error}. The statistical deviations associated to finite size effects decrease with $N$.


\begin{thebibliography}{}
\bibitem{comment} This functional
relation must nonetheless be taken in a cautious way, indeed, other chaos indicators (such as the Lyapunov spectra) may also play a relevant role in the final shape of $P(k)$ and such issues should be investigated in detail.
\end{thebibliography}

\end{document}